\documentstyle[preprint,aps,12pt]{revtex}

\begin{document}

\title{Adiabatic Effective Action for Vortices
in Neutral and Charged Superfluids}

\author{M.\ Hatsuda$^1$,
M.\ Sato$^2$, S.\ Yahikozawa$^2$
and T.\ Hatsuda$^3$
}

\address{$^1$National Laboratory for High Energy Physics,
Oho 1-1, Tsukuba, Ibaraki 305, Japan}

\address{$^2$ Physics Department, Kyoto University,
Kyoto 606-01, Japan}

\address{$^3$ Institute of Physics, University of Tsukuba,
 Tsukuba, Ibaraki 305, Japan}


\maketitle

\begin{abstract}

Adiabatic effective action for vortices in neutral and charged superfluids
 at zero temperature
 are calculated using the topological Landau-Ginzburg theory
 recently proposed by Hatsuda, Yahikozawa, Ao and Thouless,
and vortex dynamics are examined.
The Berry phase term arising in the effective action
 naturally yields  the Magnus force in both
 neutral and charged superfluids.
 It is shown that in neutral superfluid there is only one degree of freedom,
namely the center of  vorticities, 
and the vortex energy is proportinal to the sum of all vorticities
so that it is finite only for the vanishing total vorticity of the system.
 On the other hand the effective mass and the vortex energy for a vortex in 
charged superfluids are defined individually as expected.
The effects of the vortex core on
 these quantities are also estimated.
The possible depinning scenario
which is governed by the Magnus force and the inertial mass 
is also discussed.

\end{abstract}

\newpage
\setcounter{page}{1}

\parskip=7pt

\section{ Introduction}

Vortex dynamics play an essential role in  various
 properties of superfluids and superconductors\cite{Kim,Till}.
 Recently,   numbers of experiments
  on the Hall effect\cite{JnOn}
 and the anomalous Hall effect\cite{WnTn} in superconductors 
  have provided renewed interest in the theoretical study of 
vortex dynamics 
\cite{AoThMag,AT,NAT,TLG,Gai,Klee,Zee,SY,Kura}. 
 A standard theory of vortex dynamics has, however,  not
 been established yet.
 For example, the existence of the Magnus force in superconductors 
is still an open question. In fact,  
 Galilean invariance, which is used to derive the Magnus force 
in neutral superfluids,
is unclear in charged superfluids because of the presence of
a metallic lattice
 \cite{Till}.

It is worth mentioning here that 
  classical hydrodynamics  \cite{Lamb} 
provides an 
 explanation of  two main properties of vortices
in a perfect fluid; (i) Helmholtz's theorem 
and Kelvin's theorem lead to a vortex motion 
 such that a vortex-line moves with  
the background fluid,
 and (ii) Bernoulli's theorem leads to the Magnus force
acting on a vortex-line at rest.
  The nonlinear 
 Schr$\ddot{\rm o}$dinger approach \cite{GPita},
 which is one of the successful quantum descriptions
   for superconductors and superfluids, 
    contains similar hydrodynamic equations:
 When the wave function is decomposed into the phase and
the amplitude, $\psi=\sqrt{\rho}e^{\theta}$, the field equation
for $\theta$ becomes  the continuity equation and the field equation 
for $\rho$ gives  Bernoulli's theorem.
 Since this theory contains Bernoulli's theorem,
it should lead to the Magnus force.
 In fact, there exists  an attempt
to calculate the expectation value of the Hall (transverse)
 conductivity and the longitudinal conductivity 
due to vortex motions by solving the time-dependent nonlinear 
 Schr$\ddot{\rm o}$dinger equation numerically \cite{Dor}.
 However, it is a non-trivial task to extract
  the qualitative features of 
 vortex dynamics, in particular the Magnus force, from 
 such simulations.

In this paper, we develop an alternative approach to
 vortex dynamics, 
 ``the topological Landau-Ginzburg theory" which
 was proposed in a previous paper by two of us with
 Ao and Thouless \cite{TLG}.
 This is a field theoretical approach whose 
 field equations include the conventional 
nonlinear 
 Schr$\ddot{\rm o}$dinger equation.
 In this approach, by using  
 Helmholtz's and Kelvin's theorems  as guiding principles,
  topological terms and collective coordinates for 
 the vortex center  are introduced.
The hydrodynamical 
vortex law, ``$\dot{\mbox{\boldmath $X$}} =
{\mbox{\boldmath $v$}}({\mbox{\boldmath $X$}})$", is naturally
 obtained by taking a variation of our ``topological action". 
 Furthermore, the vortex motion can be
  extracted even after taking into account the 
phonon and  photon effects.

We will show  that
the topological term,
 ``$\epsilon^{\mu\nu\rho\lambda}b_{\mu\nu}f_{\rho\lambda}$",
and the source term in our action give rise to the Magnus force. 
This is consistent  with the recent work by Ao and Thouless
in which it is shown that the Berry phase and the dynamical
phase in the quantum mechanical description of  vortex system 
give rise to
the Magnus force \cite{AoThMag}. 
 In our approach, 
the Magnus force appears naturally in both 
 superconductors and  superfluids,
  which is consistent with the  
  expectation before \cite{AoThMag}
  and  with the recent results 
  \cite{Gai,Klee} in different approaches. 

Our topological action also provides us 
 with  vortex-phonon couplings, where 
 the phonon here is the density fluctuation of the
 order parameter
  at zero temperature,
namely the zero sound wave.
 In actual superfluids and superconductors,
  there are situations where the vortices are trapped
   by pinning potentials instead of moving with the
    background flow. If the background flow becomes
 strong enough,   depinning occurs 
 and  vortices  start to move.
 A candidate of the driving force to overcome the pinning
 at zero temperature is
   the Magnus force  \cite{Kim,Till}, and  the dynamics of
    depinning is dictated by the inertial mass
     of a vortex. An alternative candidate is 
quantum tunneling \cite{tunnel},
 but  we do not consider it in this paper.
 We will calculate the inertial mass as well as the vortex energy
   in  neutral and charged superfluids
  by taking into account 
 the phonon and photon contributions.
  Also evaluated are the core contributions to the inertial mass
and the vortex energy in charged superfluids where the size of 
a vortex-core is generally larger than the atomic scale.
 The obtained inertial mass is applicable to a wider
 range of the parameter region than that obtained in
  the phenomenological time-dependent 
 Landau-Ginzburg theory supplemented with the Fermi liquid
 theory \cite{DLg}.

 The organization of this paper is as follows.
 In section II, a close analogy between the 
  Lorentz force in electrodynamics and the
   Magnus force in hydrodynamics is discussed
    using our topological action for the superfluid 
     without vortex-phonon interaction. 
     In section III, the effective vortex dynamics 
in a superfluid are studied by taking into account 
the phonon interaction. In particular,  
 the inertial mass  and vortex energy 
   are calculated.
 In section IV, vortex dynamics in a superconductor
are examined by taking into account the photon interaction
and the density fluctuations. The inertial mass and vortex energy  
 in conventional and high $T_c$ 
  superconductors are then evaluated.
 Section V is devoted to summary and concluding remarks. 
 
\section{The Magnus Force}

In this section, by using the
 topological Landau-Ginzburg theory without phonon/photon
 fluctuations, 
  we will demonstrate that the origin of the Magnus force
 in superfluids 
 is quite analogous to that of the Lorentz force in 
 electrodynamics.

 The action of topological Landau-Ginzburg theory for the 
 neutral superfluid reads
\begin{eqnarray}
&&S=\int\! d^4x \biggl[\psi^*\left(i\hbar\partial_0 +
\hbar a_0 \right)\psi
- \frac{1}{2m}\left\vert\left(i\hbar\partial_i+
 \hbar a_i 
\right)
\psi\right\vert^2
- g\left(\vert \psi \vert^2-\rho_0\right)^2\nonumber \\
 & & \qquad\qquad 
 +\frac{\hbar}{2m}\varepsilon^{\mu\nu\rho\sigma}b_{\mu\nu}
f_{\rho\sigma} + b_{\mu\nu}J^{\mu\nu}\biggr] -U_{pin}(X) ,
\label{TLGSF}
\end{eqnarray}
\begin{eqnarray}
J^{\mu\nu}(x)=\sum_{a=1}^{N}\gamma_a \int\!d\tau  d\sigma 
\frac{\partial X_a^{[\mu}}
{\partial \tau} \frac{\partial X_a^{\nu ]}}{\partial \sigma}
  \delta^{(4)}
\left(x-X_a(\sigma, \tau)\right) \qquad \ \ ,
\label{TLGSF1}
\end{eqnarray}
where $\gamma_a = 2\pi n_a \hbar/m$ with integer $n_a$
and 
 $ 2\pi\hbar/m$
 denotes vorticity unit. 
 Notations follow from our previous paper \cite{TLG}.
$a_{\mu}$ is a vector potential representing vortex singularities,
$b_{\mu\nu}$ is a rank-two anti-symmetric tensor potential,
  $J^{\mu\nu}$
is the vorticity current, and $U_{pin}(X)$ is a pinning potential.
We examine vortex dynamics by taking a variation 
of $S$ with respect to
the vortex coordinate $ X$.

There is a one-to-one correspondence of our theory 
 to electrodynamics where  a point-like charged particle has a current
\begin{eqnarray}
J^{\mu} (x)=e\int d\tau \frac{dX^{\mu}}{d\tau}
\delta^{(4)}(x-X(\tau)) .
\end{eqnarray}
Variation of the source term $-\int d^4x A_{\mu}J^{\mu}$
 in electrodynamics with respect to the particle
coordinate $X(t)$ gives the Lorentz force 
\begin{eqnarray}
-\frac{\delta}{\delta X^i (t)}\int d^4x A_{\mu} J^{\mu} 
& = & -eF_{i\mu}(X) \frac{dX^{\mu}(t)}{dt}\label{Lrz}
 \\ \nonumber
& = & e({\mbox{\boldmath $E$}}(X) 
+\dot{\mbox{\boldmath $X$}}\times{\mbox{\boldmath $B$}}(X) )^i 
\\ \nonumber
&=&F_{Lorentz}^i(X)\ \ \ ,
\end{eqnarray}
where the proper-time variable $\tau$ is chosen to be $\tau=t=X^0$.

Analogously  a variation of the source term 
$\int d^4x   bJ $ in (\ref{TLGSF}) with respect to the vortex
  coordinate gives
\begin{eqnarray}
\frac{\delta}{\delta X^i (t,\sigma)}\int d^4x  b_{\mu\nu} J^{\mu\nu}
=\gamma H_{i\mu\nu}(X) \frac{\partial X^{[\mu}}{\partial t}
\frac{\partial X^{\nu ]}}{\partial\sigma}
\ \ \ ,
\label{dbj1}
\end{eqnarray}
where   
$H_{\mu\nu\rho}=\partial_{\mu}b_{\nu\rho}+\partial_{\nu}b_{\rho\mu}
+\partial_{\rho}b_{\mu\nu}$ and 
 $\tau =t=X^0$. On the other hand, 
the field equation for $a_{\mu}$ obtained from 
 (\ref{TLGSF}) gives a relation between the 
$b$-field 
and the hydrodynamical current, 
${\cal J}^{\mu}(x) = (m\rho(x) , m\rho(x) v^i(x) ) $;
\begin{eqnarray}
H_{\mu\nu\rho} = -\frac{1}{2}\varepsilon_{\mu\nu\rho\lambda} 
{\cal J}^{\lambda}
\quad .
\label{HeqJ} 
\end{eqnarray}
The $b$-field is the antisymmetric rank-two tensor potential
whose exterior derivative is the observable hydrodynamical current.
This property is analogous to the vector potential
 whose exterior derivative
 is  electromagnetic field.
 Note that 
 $H_{i\mu\nu}(X)$ in eq.(\ref{dbj1}) should be considered as an 
 external field acting on the vortex at point $X$, 
 which is analogous to  
the electromagnetic field acting on the charged particle in eq.(\ref{Lrz}).
Inserting eq.(\ref{HeqJ}) into eq.(\ref{dbj1}) leads to the conventional
 Magnus force
\begin{eqnarray}
\frac{\delta}{\delta X^i (t,\sigma)}\int b_{\mu\nu} J^{\mu\nu} dx
&=&-\varepsilon_{i\mu\nu\rho} {\cal J}^{\rho} (X)
\frac{\partial X^{\mu}}{\partial t}
\left( \gamma \frac{\partial X^{\nu }}{\partial \sigma}\right)
 \nonumber \\
 &=&- m\rho(X) [(\dot{\mbox{\boldmath $X$}}-
{\mbox{\boldmath $v$}}(X))\times{\mbox{\boldmath $\omega$}} (X)]^i
\nonumber\\
 &=&F^i_{Magnus}(X),
\quad 
\label{mag1}
\end{eqnarray} 
where 
we have used the fact that
$\left( \gamma \frac{\partial {\mbox{\boldmath $X$}}}
{\partial \sigma}\right) $
is  the vorticity vector, ${\mbox{\boldmath $\omega$}}\equiv
 \nabla\times{\mbox{\boldmath $v$}}$.
 The Magnus force, which is analogous to
the Lorentz force in electrodynamics,
  is thus the fundamental force in the hydrodynamic theory.
 
If there are no pinning  potentials, the variation of $S$ 
at the 
``classical" level
 $\partial S/ \partial X^i(t,\sigma)=0$ gives an equation
of a vortex
 $F^i_{Magnus}(X)= -m\rho(X)
 [(\dot{\mbox{\boldmath $X$}}-
{\mbox{\boldmath $v$}}(X))\times
{\mbox{\boldmath $\omega$}} (X)]^i=0$.
 Analogous to the case of the electrodynamics,
  ${\cal J}^{\mu}(X)$ and $H^{\mu\nu\rho}(X)$ 
 take values at point
  $X$ without the contribution of a vortex sitting at $X$ in
 the weak current limit;
  namely $\rho (X) \neq 0$ and ${\mbox{\boldmath $v$}}(X) \neq 0$.
 Thus one arrives at the conventional classical
  hydrodynamical relation
 $\dot{\mbox{\boldmath $X$}}=
{\mbox{\boldmath $v$}}({\mbox{\boldmath $X$}})$ 
which means  that the vortex moves
 with the same velocity as the background flow.
 Even if one takes into account the full vortex contributions to 
 ${\cal J}^{\mu}(X)$, as far as one considers the finite size of 
 a vortex core, it is possible to derive similar hydrodynamical
 relation in an averaged sense:
\begin{eqnarray}
\langle\dot{\mbox{\boldmath $X$}}\rangle
 =\frac{\langle\rho{\mbox{\boldmath $v$}}({\mbox{\boldmath $X$}})\rangle}
{\langle\rho({\mbox{\boldmath $X$}})\rangle}
\quad .\label{xdv}
\end{eqnarray}
 In fact, if one uses a distribution with finite 
core size instead of the  $\delta$-function
 distribution given in (\ref{TLGSF1}),
 the vortex centers $X_a$ and the current ${\cal J}^{\mu}(X)$
 must be defined by averaging over the distribution as
 denoted in by $\langle \cdot \rangle$ in (\ref{xdv}).
For ``classical hydrodynamical'' cases
 (\ref{xdv}) further reduces to $\langle \dot{\mbox{\boldmath $X$}} \rangle 
= \langle {\mbox{\boldmath $v$}}({\mbox{\boldmath $X$}}) \rangle $
 since $\langle\rho{\mbox{\boldmath $v$}}({\mbox{\boldmath $X$}})\rangle = 
 \langle\rho({\mbox{\boldmath $X$}})\rangle 
\langle{\mbox{\boldmath $v$}}({\mbox{\boldmath $X$}})\rangle$ 
 holds.
One can imagine easily this situation
for a case such as an isolated straight vortex line along the
 $z$-direction and a constant background flows in the x-y plane.

The above discussions  are valid not only for  neutral superfluids
but also for  charged superfluids since the 
$bJ$ coupling is universal in both cases. Namely
 the Magnus force is a fundamental 
force in both neutral and charged superfluids.

Up to this point, we have not taken into account the interactions
 of phonons or photons with vortices. They actually induce a
 kinietic term ${1 \over 2} m_{eff} \dot{X}^2$ with
 $m_{eff}$ being the inertial mass.  This means that,
 beyond the ``classical'' level of $S$,
the acceleration
 term is induced from the equation of motion of $X$, i.e.,
 $m_{eff} \dot{X}^2 = F_{Magnus}(X) + F_{pin}(X)$. 
 This will be discussed in detail  
 in later sections.

\section{effective action 
for vortices in  neutral superfluid }

Now let us calculate the effective action for vortices in 
a neutral superfluid by taking into account
the phonon interaction at zero temperature.
We begin with the topological Landau-Ginzburg
theory (\ref{TLGSF}) and (\ref{TLGSF1})
 and define the phase and amplitude variables
 as $\psi =\sqrt{\rho (x)} e^{i\theta (x)}$. Thus the action $S$ reads
\begin{eqnarray}
S=\int\! d^4x \biggl[\hbar\rho \left(-\partial_0\theta +a_0 \right)
- \frac{\hbar^2\rho}{2m}\left(\nabla\theta +{\mbox{\boldmath $a$}}
\right)^2 
-\frac{\hbar^2}{8m\rho}(\nabla \rho )^2
-g(\rho -\rho_0)^2
\nonumber \\
 +\frac{\hbar}{2m}\varepsilon^{\mu\nu\rho\sigma}b_{\mu\nu}
f_{\rho\sigma} + b_{\mu\nu}J^{\mu\nu}\biggr] -U_{pin}(X)
\quad\quad .
\label{TLGf1}
\end{eqnarray}
 If there is a uniform background flow
 ${\mbox{\boldmath $v$}}_{bg}=$constant and vortex excitations do not exist,
 the stationary state of the superfluid is characterized as
\begin{eqnarray}
a^{\mu} =0\quad ,\quad \rho=\rho_0 \quad ,\quad 
\theta =\theta_{bg} \equiv 
\frac{m}{\hbar}{\mbox{\boldmath $v$}}_{bg} \cdot {\mbox{\boldmath $r$}}
-\frac{m}{2\hbar}{\mbox{\boldmath $v$}}_{bg} ^2t \quad\quad .
\label{bg}
\end{eqnarray}
In order to take into account the phonon fluctuation, we expand 
 the amplitude and phase variables around the
 above stationary solution, $\rho(x)=\rho_0+\delta \rho$,
 $\theta(x) = \theta_{bg} + \delta \theta$.
 Then the full functional integral and the effective action
 $S_{eff}(X)$ can be written as
\begin{eqnarray}
{\cal Z}=\int {\cal D}[\delta\rho , \delta\theta ,a_{\mu},  b_{\mu\nu}]
e^{iS(\rho , \theta , a_{\mu},  b_{\mu\nu};X)}
=e^{iS_{eff} (X)},
\label{ef}
\end{eqnarray}
where $S (\rho , \theta , a_{\mu},  b_{\mu\nu};X) = \int d^4x {\cal L} 
 - U_{pin}(X)
$ with
\begin{eqnarray}
 {\cal L}&=&{\cal L}_{st} 
+\delta\rho \left( \hbar a_0 -
\hbar{\mbox{\boldmath $v$}}_{bg}\cdot{\mbox{\boldmath $a$}}
-\frac{\hbar^2}{2m}{\mbox{\boldmath $a$}}^2 \right)
+\frac{1}{2}(\delta\rho\quad \delta\theta)G^{-1}
\left( \begin{array}{c} 
        \delta\rho\\ \delta\theta\end{array} \right)
\nonumber\\ &&\quad\quad
-\frac{\hbar^2}{m}\delta\rho{\mbox{\boldmath $a$}}
\cdot\nabla\delta\theta
+O(\delta^3)
+ \left( bf,\ bJ \right) 
\nonumber , \\
{\cal L}_{st} &=&
\rho_0 \left( \hbar a_0 -\hbar{\mbox{\boldmath $v$}}_{bg}
\cdot{\mbox{\boldmath $a$}}
-\frac{\hbar^2}{2m}{\mbox{\boldmath $a$}}^2 \right).
\label{TLGf2}
\end{eqnarray}
Here ${\cal L}_{st}$ is a  stationary part of the action and
 $G^{-1}$ is a Hermitian matrix defined as
\begin{eqnarray}
G^{-1}=\left(\begin{array}{cc}
        \left( -2g+\frac{\hbar^2}{4m\rho_0}\triangle \right)& 
        \left( -\hbar\partial_0-
        \hbar{\mbox{\boldmath $v$}}_{bg}
        \cdot\nabla \right)\\
         & \\
        \left(\hbar\partial_0 
        +\hbar{\mbox{\boldmath $v$}}_{bg}
        \cdot\nabla \right)
        & \frac{\rho_0\hbar^2}{m}\triangle
                \end{array}\right)
                \quad\quad ,
\end{eqnarray}            
where we used the Coulomb gauge for the $a$-field and 
divergence free property of the background flow.
 $G$ is nothing but the Green's function for the fluctuations
 and has poles at
\begin{eqnarray}
\omega = {\mbox{\boldmath $v$}}_{bg}\cdot{\mbox{\boldmath $k$}}
        \pm \epsilon ({\mbox{\boldmath $k$}}),
        \quad  \quad
{\rm with} \ \ \ \ \epsilon^2 ({\mbox{\boldmath $k$}}) 
=c_s^2{\mbox{\boldmath $k$}}^2
+\left( \frac{{\mbox{\boldmath $k$}}^2}{2m}\right)^2,
\label{Bog}
\end{eqnarray}
with $\epsilon ({\mbox{\boldmath $k$}}) $ being  the Bogoliubov spectrum       
and $c_s$ being  the zero sound velocity
\begin{eqnarray}
c_s =\sqrt{\frac{2\rho_0 g}{m}} \quad  .
\label{cs}
\end{eqnarray}
For small momentum ${\mbox{\boldmath $k$}}$, 
this reduces to the massless phonon mode
propagating with sound velocity under the background flow.
After the Gaussian integration with respect to $\delta\rho$
and $\delta\theta$, the effective lagrangian
 becomes
\begin{eqnarray}
&& {\cal L}_{eff}(a_{\mu},b_{\mu\nu};X)\nonumber\\
&&\quad\quad\quad\quad\approx 
 {\cal L}_{st}
 + \frac{\rho_0\hbar^2}{2m}  
\left( a_0
- {\mbox{\boldmath $v$}}_{bg}\cdot{\mbox{\boldmath $a$}}
-\frac{\hbar}{2m}{\mbox{\boldmath $a$}}^2 
\right) 
\left( \frac{-\triangle}{(\partial_0
+{\mbox{\boldmath $v$}}_{bg}\cdot\nabla)^2+\epsilon^2}\right)
\left( a_0
- {\mbox{\boldmath $v$}}_{bg}\cdot{\mbox{\boldmath $a$}}
-\frac{\hbar}{2m}{\mbox{\boldmath $a$}}^2 
\right)
\nonumber\\&&\quad\quad\quad\quad\quad
 +\frac{\hbar}{2m}\varepsilon^{\mu\nu\rho\sigma}b_{\mu\nu}
f_{\rho\sigma} + b_{\mu\nu}J^{\mu\nu}
\nonumber\\
&&\quad\quad\quad\quad\approx
 {\cal L}_{st}
 + \frac{\rho_0\hbar^2}{2m} \frac{1}{c_s^2} 
\left( a_0
- {\mbox{\boldmath $v$}}_{bg}\cdot{\mbox{\boldmath $a$}}
-\frac{\hbar}{2m}{\mbox{\boldmath $a$}}^2 
\right)^2 
 +\frac{\hbar}{2m}\varepsilon^{\mu\nu\rho\sigma}b_{\mu\nu}
f_{\rho\sigma} + b_{\mu\nu}J^{\mu\nu}
\quad\quad .
\nonumber \\&&
\label{TLGf3}
\end{eqnarray}
For interactions between vortices, 
only the phonon modes with small momentum are important
under normal circumstances, 
since the vortex motion and the background flow are 
much slower than the zero sound velocity;
$\dot{X}<<c_s$, $v_{bg}<<c_s$.
 Therefore, from the second equality to the last one in (\ref{TLGf3}),
$(\partial_0+{\mbox{\boldmath $v$}}_{bg}\cdot\nabla)^2$ is neglected
and $\epsilon^2\approx -c_s^2\triangle$ is taken 
 (the adiabatic approximation).

The integration of 
the $b_{\mu\nu}$ field is straightforward and the following constraint
on $a_{\mu}$ is obtained: 
\begin{eqnarray}
f_{\rho\lambda}=\partial_{\rho}a_{\lambda} - \partial_{\lambda}a_{\rho} 
=\frac{m}{2\hbar}
\varepsilon_{\rho\lambda\mu\nu} J^{\mu\nu}
\quad\quad .
\label{fdaJ}
\end{eqnarray}
This constraint (\ref{fdaJ}) can be solved in the Coulomb gauge as  
\begin{eqnarray}
\left\{ \begin{array}{ccl}
a_0&=&\frac{-1}{\triangle} (m/2\hbar) 
\epsilon_{ijk} \partial^i J^{jk} \\
 &&\\
a_i&=&\frac{-1}{\triangle} (m/\hbar) 
\epsilon_{ijk} \partial^j J^{0k} 
        \end{array} \right.
\quad\quad .
\end{eqnarray} 
In the following, we will mostly focus our attention on the
 vortices in two spatial dimensions where
 vortices lie along the z-direction to make the argument as simple as 
 possible.  In this case,
  the $a_{\mu}$ field
 is written as
\begin{eqnarray}
\left\{\begin{array}{l}
a_0=
-\sum_a n_a {\mbox{\boldmath $e$}}_z \cdot 
\left(\dot{\mbox{\boldmath $X$}}_a\times\nabla\ln|{\mbox{\boldmath $x$}}
-{\mbox{\boldmath $X$}}_a| \right)
\\
{\mbox{\boldmath $a$}}=
\sum_a n_a {\mbox{\boldmath $e$}}_z 
\times\nabla\ln |{\mbox{\boldmath $x$}}-{\mbox{\boldmath $X$}}_a| 
\end{array}\right.
\quad  .
\label{hata}
\end{eqnarray}

 We will now insert the solutions (\ref{hata}) into
 (\ref{TLGf3}) to get an effective action $S_{eff}(X)$ written
 in the vortex coordinate alone.
 When doing this,
 it is convenient to classify terms by the number of $a_{\mu}$-fields.
The $a_{\mu}$ field contains the vortex singularities,
so the number of $a_{\mu}$ 
 fields represents the number of the interacting vortices;
for example, a bilinear term of  $a_{\mu}$
 field contains an interaction of two vortices.
Let us first evaluate the term linear in $a_{\mu}$ field.  
Inserting the expressions (\ref{hata}) into the linear term
 in (\ref{TLGf3}), one gets
 \begin{eqnarray}
\int d^2x\rho_0\hbar ({a}_0-{\mbox{\boldmath $v$}}_{bg}
\cdot{\mbox{\boldmath $a$}}) 
&=&-\int d^2x \rho_0\hbar\sum_a n_a{\mbox{\boldmath $e$}}_z
\cdot(\dot{\mbox{\boldmath $X$}}_a -{\mbox{\boldmath $v$}}_{bg})
\times\nabla \ln|{\mbox{\boldmath $x$}}-{\mbox{\boldmath $X$}}_a|
\nonumber \\ \label{Bpm1}
 & = &\frac{m\rho_0}{2}\sum_a \gamma_a
{\mbox{\boldmath $e$}}_z
\cdot(\dot{\mbox{\boldmath $X$}}_a -
{\mbox{\boldmath $v$}}_{bg})\times({\mbox{\boldmath $X$}}_a-{\mbox{\boldmath $v$}}_{bg}t)
\quad \quad ,
\label{Bpmm1}
\end{eqnarray}
 where one should be careful 
to perform spatial integration under the 
background flow. In fact,
 there is a constant ambiguity in the integral, 
$\int d^2x \nabla \ln |{\mbox{\boldmath $x$}}
-{\mbox{\boldmath $X$}}|=
-\pi({\mbox{\boldmath $X$}}+{\mbox{\boldmath $c$}})$.
  ${\mbox{\boldmath $c$}}$ can be fixed as 
  $-{\mbox{\boldmath $v$}}_{bg}t$
 by requiring a correct boundary condition,
 i.e.,  the boundary of the superfluid should also move
   with the background flow velocity.
 From (\ref{Bpmm1}), one sees that
 the combination ${a}_0-
{\mbox{\boldmath $v$}}_{bg}\cdot{\mbox{\boldmath $a$}}$
 has  the Galilean invariance.

The obtained expression (\ref{Bpmm1}) is a generalization
 of the Berry phase term \cite{berry-p} to the system with
 background flow. 
 It is easy to show that 
taking a variation of this term 
 with respect to the vortex coordinates leads to
 the Magnus force.

Secondly, let us evaluate bilinear terms in the $a_{\mu}$ field.
In order to evaluate them we will use the following formula:
\begin{eqnarray}
&&\int d^2x \partial_i
\ln|{\mbox{\boldmath $x$}}-{\mbox{\boldmath $X$}}_a|
\partial_j
 \ln|{\mbox{\boldmath $x$}}-{\mbox{\boldmath $X$}}_b| 
\nonumber\\
&&\quad\quad=
\left\{ \begin{array}{lcl}
-\pi\{\delta^{ij}\ln({|{\mbox{\boldmath $X$}}_{ab}|}/{R})+
({X_{ab}^i X_{ab}^j}/{{\mbox{\boldmath $X$}}_{ab}^2}-\delta^{ij}/{2})\}
&,& {\rm for}{\mbox{\boldmath $X$}}_a\neq {\mbox{\boldmath $X$}}_b\\
-\pi\delta^{ij}\ln({d}/{R})
&,& {\rm for}{\mbox{\boldmath $X$}}_a= {\mbox{\boldmath $X$}}_b
\end{array}\right.
\quad\quad ,
\label{rlt}
\end{eqnarray}
where $d$ is an ultraviolet cutoff which is an atomic scale,
  $R$ is an infrared cutoff and 
${\mbox{\boldmath $X$}}_{ab}
\equiv{\mbox{\boldmath $X$}}_{a}-{\mbox{\boldmath $X$}}_{b}$.
Appendix A contains a derivation of the above formula 
by regulating $\ln |{\mbox{\boldmath $x$}}|$
 by $\ln\sqrt{{\mbox{\boldmath $x$}}^2
+\varepsilon^2}$ with $d=\varepsilon e^{1/2}$.
It is natural to choose $R$ to be the container size.  
 In a charged superfluid,  the penetration depth
 plays the role of the infrared cutoff $R$,
 since the charge screening makes 
 the interaction region finite.

Using the formula (\ref{rlt}), $\int {\mbox{\boldmath $a$}}^2$ term in 
(\ref{TLGf2}) and (\ref{TLGf3}) becomes
\begin{eqnarray}
-\frac{\rho_0\hbar^2}{2m}\int d^2x {\mbox{\boldmath $a$}}^2
 &=& -\frac{\rho_0\hbar^2}{2m} (\sum_{a=b}+\sum_{a\neq b})n_a n_b
\int d^2x \ \nabla\ln|{\mbox{\boldmath $x$}}
-{\mbox{\boldmath $X$}}_a| \cdot\nabla  
\ln|{\mbox{\boldmath $x$}}-{\mbox{\boldmath $X$}}_b| 
\nonumber\\
 &= & 
 -\frac{\rho_0\hbar^2\pi}{m}\{ \sum_{a} n_a^2
  \ln\frac{R}{d}
+ \sum_{a\neq b} n_a n_b
 ( \ln\frac{R}{d} -\ln\frac{|{\mbox{\boldmath $X$}}_{ab}|}{d})\}\nonumber\\
&=&-E_0+  
 \sum_{a\neq b}\frac{m\rho_0}{4\pi} \gamma_a \gamma_b
  \ln\frac{|{\mbox{\boldmath $X$}}_{ab}|}{d}  
\label{intm3}
\quad\quad .
\end{eqnarray}
Here $E_0$ is the static energy for a system
 with many vortices,
\begin{eqnarray}
 E_{0}=\frac{m\rho_{0}}{4\pi}(\sum_a \gamma_a)^2\ln\frac{R}{d}\ \ .
\label{e00}
\end{eqnarray}
 The obtained expression  $E_0$
for an isolated vortex coincides with Feynman's result 
\cite{Fynm}. 
If the container size is infinite,  $E_0$
 for a single vortex is divergent, 
 while $E_0$ vanishes for a system with zero total vorticity.
  For example, a vortex $-$ anti-vortex pair has
 zero static energy and has only the logarithmic
 interaction  in (\ref{intm3}).

Among the self-interaction terms, 
there is a term which is interpreted as the kinetic term;
\begin{eqnarray}
&&\int d^2x \frac{\rho_0\hbar^2}{2mc_s ^2} 
\left({a}_0-{\mbox{\boldmath $v$}}_{bg}\cdot{\mbox{\boldmath $a$}}\right)^2
\nonumber\\&  & \quad\quad
 = 
\frac{\rho_0\hbar^2}{2mc_s ^2}( \sum_{a=b}+\sum_{a\neq b})n_an_b 
 (\dot{\mbox{\boldmath $X$}}_a-
{\mbox{\boldmath $v$}}_{bg})^i
(\dot{\mbox{\boldmath $X$}}_b-{\mbox{\boldmath $v$}}_{bg})^j
\epsilon^{ik}\epsilon^{jl}
\int d^2x \partial_k\ln|{\mbox{\boldmath $x$}}
-{\mbox{\boldmath $X$}}_a|\partial_l
\ln|{\mbox{\boldmath $x$}}-{\mbox{\boldmath $X$}}_b|
\nonumber \\
 & & \quad\quad
  =
\frac{1}{2}m_{eff}
(\sum_a {\gamma_a \over \gamma_0}
(\dot{\mbox{\boldmath $X$}}_a-{\mbox{\boldmath $v$}}_{bg}))^2
\nonumber\\&&
 \quad\quad
 -\frac{m\rho_0}{8\pi c_s ^2}\sum_{a\neq b} \gamma_a \gamma_b
\{
(\dot{\mbox{\boldmath $X$}}_a-
{\mbox{\boldmath $v$}}_{bg})\cdot
(\dot{\mbox{\boldmath $X$}}_b-{\mbox{\boldmath $v$}}_{bg})
(\ln\frac{|{\mbox{\boldmath $X$}}_{ab}|}{d}+\frac{1}{2})
-(\dot{\mbox{\boldmath $X$}}_a-
{\mbox{\boldmath $v$}}_{bg})\cdot {\mbox{\boldmath $u$}}_{ab}
(\dot{\mbox{\boldmath $X$}}_b-
{\mbox{\boldmath $v$}}_{bg})\cdot {\mbox{\boldmath $u$}}_{ab}
\}
\nonumber\\
\label{int2}
\end{eqnarray}
where 
${\mbox{\boldmath $u$}}_{ab}
\equiv{\mbox{\boldmath $X$}}_{ab}/|{\mbox{\boldmath $X$}}_{ab}|$. 
Here we define
\begin{eqnarray}
{\mbox{\boldmath $X$}}_{CV}\equiv \sum_a{\gamma_a \over \gamma_0}
({\mbox{\boldmath $X$}}_a-{\mbox{\boldmath $v$}}_{bg}t)
\quad ,\quad
\gamma_0\equiv \frac{2\pi\hbar}{m}
\label{cm}
\end{eqnarray}
and call this ``the center of vorticity'' of the system.
This is analogous to the center of mass
except that  vorticity takes
both positive and negative value. 
Then the first term of eq.(\ref{int2}) 
 is interpreted as a  kinetic term for ${\mbox{\boldmath $X$}}_{CV}$ 
with the following inertial mass
\begin{eqnarray}
m_{eff}=\frac{m\rho_0}{4\pi c_s ^2}
\gamma_0^2\ln\frac{R}{d} 
\quad\quad .
\label{meff}
\end{eqnarray}
If phonon fluctuations are neglected (i.e., in the limit
  $c_s\rightarrow 0$), 
``the center of vorticity'' 
$\sum \gamma_a {\mbox{\boldmath $X$}}_a/\gamma_0$
 is conserved  since it is proportional to the
 conserved total momentum
 ${\mbox{\boldmath $P$}}=
\sum {\mbox{\boldmath $p$}}_a=
-(m\rho_0/2){\mbox{\boldmath $e$}}_z\times
\sum \gamma_a{\mbox{\boldmath $X$}}_a$.
 If there exist  phonons, 
${\mbox{\boldmath $P$}}$ has an extra term from the phonon
 momentum and  $\sum \gamma_a {\mbox{\boldmath $X$}}_a$ 
is not conserved anymore.
 Instead, it gets mobility with an inertial mass $m_{eff}$ given
 in (\ref{meff}).

The magnitude of the vortex energy $E_0$ in (\ref{e00}) and the 
 inertial mass $m_{eff}$ in (\ref{meff}), are
consistent
 with those in ref.\cite{DLg} and ref.\cite{Popov}
obtained by different approaches,
although the vorticity dependence was not considered in these 
references.
Ref. \cite{DLg} is based 
on a phenomenological time-dependent Landau-Ginzburg theory 
 supplemented with the Fermi liquid theory,
 while ref. \cite{Popov}
is based on the functional integral approach at finite temperature.
It is, however,  unclear whether our result is consistent with ref.
\cite{NAT} which discusses the behavior of $m_{eff}$ for large $R$.

Lastly we evaluate the vortex interactions.
The leading term of two-vortex interaction 
is given by the second term of  the right hand side of
(\ref{intm3}).
Next to leading terms such as the second term in (\ref{int2})
are suppressed by the factor
${\mbox{\boldmath $v$}}_{bg}^2/c_s ^2$ which is of order 
$10^{-4}\sim 10^{-8}$ for superfluids He$^4$,
 $10^{-2}$ for 
conventional superconductors and $10^{-6}$ for 
high $ T_c$ superconductors. 
The suppression factor 
$\dot{\mbox{\boldmath $X$}}^2/c_s ^2$ is even smaller than
 ${\mbox{\boldmath $v$}}_{bg}^2/c_s ^2$ 
because we consider the adiabatic motion
 of  vortices.
Three and four vortex interactions
 are generated from third order and fourth order terms in $a_{\mu}$.
 However, they can be neglected in a dilute vortex system.
 Furthermore these terms are suppressed by $1/c_s ^2$ and $\hbar$.
 Third and fourth order terms also contain 
contributions to the vortex self-energy and the two-vortex
interactions, but they are suppressed by $\hbar$ compared 
to leading terms in (\ref{intm3}) which are already of $O(\hbar^2)$.

The resultant effective action up to
  $O(\hbar^2)$ and $O((1/c_s^0))$ is summarized as
\begin{eqnarray}
S_{eff}(X) =
\int dt[ {L}_{Berry}+{L}_{ve}+
{ L}_{kin}+{ L}_{int}] - U_{pin}(X),
\nonumber \\ \label{easf}
\end{eqnarray}
\begin{eqnarray}
\left\{
\begin{array}{lcl}
{ L}_{Berry}&=&
\frac{m\rho_0}{2}\sum_a
 \gamma_a
{\mbox{\boldmath $e$}}_z
\cdot(\dot{\mbox{\boldmath $X$}}_a 
-{\mbox{\boldmath $v$}}_{bg})
\times({\mbox{\boldmath $X$}}_a-{\mbox{\boldmath $v$}}_{bg}t)
\\ &&\\
{ L}_{ve }&=&
-E_{0}
\quad , \quad {\rm with} \ \ \ \ 
E_{0}=({m\rho_{0}}/{4\pi})(\sum_a\gamma_a)^2\ln({R}/{d})
\\ && \\
{ L}_{kin}
&=&
\frac{1}{2}m_{eff}\dot{\mbox{\boldmath $X$}}_{CV}^2 
\quad , \quad {\rm with} \ \ \ \ 
m_{eff}=({m\rho_0}/{4\pi c_s ^2})\gamma_0^2 \ln({R}/{d})
\\ && \\
{ L}_{int}
&=& 
\sum_{a\neq b} 
(m\rho_0 /4\pi) \gamma_a\gamma_b  
 \ln(|{\mbox{\boldmath $X$}}_{ab}|/d)    ,
 \end{array}\right.\nonumber
\end{eqnarray}
where $L_{Berry}$, 
${ L}_{ve}$, 
${ L}_{kin}$ and
${ L}_{int}$ are the Berry phase term, the vortex energy,
the vortex kinetic term and the interaction term, respectively.

The equation of motion for an isolated vortex 
with $n_a=1$ at zero temperature
is determined by
varying the effective action which is the sum of 
the acceleration term,
the Magnus force, and the pinning force; 
\begin{eqnarray}
m_{eff}\ddot{\mbox{\boldmath $X$}} 
&=& -m\rho_0 
\left(\dot{\mbox{\boldmath $X$}}-{\mbox{\boldmath $v$}}_{bg}\right)
\times{\mbox{\boldmath $\omega$}}
+{\mbox{\boldmath $F$}}_{pin}\nonumber\\
&=&{\mbox{\boldmath $F$}}_{Magnus}+{\mbox{\boldmath $F$}}_{pin}
  \quad .
\label{eqmot}
\end{eqnarray}  
This equation gives the following scenario for vortex pinning and depinning.
 Consider
a vortex pinned by the pinning potential $U_{pin}(X)$. The vortex
 feels the Magnus force as long as 
 it does not move with the background flow ${\mbox{\boldmath $v$}}_{bg}$. 
If the Magnus force is strong enough to overcome 
the pinning potential, the vortex starts to move (depinning) 
 with the accerelation dictated by the inertial mass $m_{eff}$.
As the vortex velocity gets closer to the background
velocity ${\mbox{\boldmath $v$}}_{bg}$, the Magnus force gets smaller and 
 reduces eventually to zero.
  The final stationary situation is then
described by the hydrodynamical law for a perfect fluid;
 $\dot{X}= \ v(X)$. 
 The phonon interaction is irrelevant at the final stage,
but it is essential at the depinning stage.

\section{effective action for vortices in charged superfluid}

Now let us consider vortex dynamics in  type II
superconductors. We propose the following 
topological Landau-Ginzburg
theory for vortices in  superconductors;
\begin{eqnarray}
&&S=\int\! d^4x \biggl[\psi^*\left(i\hbar\partial_0 +
\hbar a_0 -\frac{q}{c}A_0\right)\psi
- \frac{1}{2m}\left\vert\left(i\hbar\partial_i+
 \hbar a_i -\frac{q}{c}A_i
\right)
\psi\right\vert^2
- g\left(\vert \psi \vert^2-\rho_0\right)^2\nonumber \\
 & & \qquad\qquad -\rho_{lat} A_0 -\frac{1}{4}F^2
+\frac{\hbar}{2m}\varepsilon^{\mu\nu\rho\sigma}b_{\mu\nu}
f_{\rho\sigma} + b_{\mu\nu}J^{\mu\nu} \biggr]  -U_{pin}(X) ,
\label{TLGSC}
\end{eqnarray}
\begin{eqnarray}
J^{\mu\nu}(x)=\sum_{a=1}^{N}\gamma_a \int\!d\tau  d\sigma 
\frac{\partial X_a^{[\mu}}
{\partial \tau} \frac{\partial X_a^{\nu ]}}{\partial \sigma}  \delta^{(4)}
\left(x-X_a(\sigma, \tau)\right)\ \ ,
\label{eq.2}
\end{eqnarray}
where $q=2e$ and $m=2m_e$ with $e$ and $m_e$ being the 
electron's charge and 
mass,  $A_{\mu}$ being the electromagnetic potential and $F_{\mu\nu}
=\partial_{\mu}A_{\nu}-\partial_{\nu}A_{\mu}$.
$\rho_{lat}$ is the electric charge of the background
lattice. In the stationary situation without vortices,
  the total electric charge vanishes locally;
$\rho_{lat}+(q/c)\rho_0=0$.

The Meissner effect
and the quantization
of magnetic flux can be easily
 checked by  the variation of the action
(\ref{TLGSC}).
The variation with respect to $A^i$ gives 
\begin{eqnarray}
\partial^{\mu} F_{\mu i} +J_e^i =0
\quad\quad ,\quad\quad
{\mbox{\boldmath $J$}}_e =-\frac{ \rho q}{mc} 
(-\hbar\nabla \theta -\hbar {\mbox{\boldmath $a$}}
+ \frac{q}{c} {\mbox{\boldmath $A$}}) 
\quad\quad .
\label{meiss}
\end{eqnarray}
 Taking curl of (\ref{meiss}) leads to 
\begin{eqnarray} 
(\Box +\frac{1}{\lambda^2} )
{\mbox{\boldmath $B$}}=\frac{\hbar c}{\lambda^2 q} 
{\mbox{\boldmath $\omega$}},
\label{meiss2}
\end{eqnarray}
where the penetration depth $\lambda$ is defined as
\begin{eqnarray}
\lambda =\sqrt{mc^2/ \rho_0 q^2}
\quad\quad .
\end{eqnarray}
Eq. (\ref{meiss2}) represents the Meissner effect that the 
magnetic field has a penetration depth.
Next let us integrate eq. (\ref{meiss2}) in the region $D$
surrounded by a closed path.  
\begin{eqnarray}
\int_D (\lambda^2\Box +1){\mbox{\boldmath $B$}} 
\cdot d{\mbox{\boldmath $n$}} 
= \frac{c\hbar}{q}\int_D {\mbox{\boldmath $\omega$}} 
\cdot d{\mbox{\boldmath $n$}} 
 =\frac{c\hbar}{q} 2\pi n  
\equiv \phi_0 n \ \ .
\end{eqnarray}
 If we choose the integration region
$D$ larger than the penetration depth around the vortex,
 $\Box {\mbox{\boldmath $B$}}$  can be  neglected because
there are neither a magnetic field nor an electric current.
Therefore the whole magnetic flux coincides with the integer multiple of
 unit magnetic flux $\phi_0=(c\hbar/q) 2\pi $.

 The Hall (or transverse) voltage and  
the longitudinal voltage are produced depending on 
the direction of the vortex motion, which follow from
the Maxwell's equations:
If a magnetic flux moves, the magnetic field changes in time
only through the flux motion and an electric field is produced by  
\begin{eqnarray}
 & &\nabla \times {\mbox{\boldmath $E$}}(x) =
-\frac{1}{c}\dot{\mbox{\boldmath $B$}}(x) 
= \frac{1}{c}(\dot{\mbox{\boldmath $X$}} \cdot \nabla )
{\mbox{\boldmath $B$}}(x)    
=-\frac{1}{c}\nabla \times (\dot{\mbox{\boldmath $X$}} 
\times {\mbox{\boldmath $B$}}(x) )\nonumber \\
 & &\quad \quad \quad \quad \quad \quad \quad 
 \rightarrow {\mbox{\boldmath $E$}}=-\frac{1}{c}
\dot{\mbox{\boldmath $X$}}\times{\mbox{\boldmath $B$}}
\quad .
\end{eqnarray}

Now let us examine vortex dynamics 
in a superconductor by taking the variation 
of our effective action (\ref{TLGSC}) of the ``classical'' level. 
Since ${\mbox{\boldmath $X$}}$ 
dependence arises only through the source term
as with the superfluid vortex system,
 the obtained force acting on vortices
in a superconductor is also the Magnus force (\ref{mag1}).
The Magnus force (\ref{mag1}) is rewritten in terms of
the magnetic field through eq.(\ref{meiss2})
\begin{eqnarray}
\label{magL}
- m\rho_0 (\dot{\mbox{\boldmath $X$}}
-{\mbox{\boldmath $v$}})\times{\mbox{\boldmath $\omega$}}
=\frac{q}{c} \rho_0
({\mbox{\boldmath $v$}}-\dot{\mbox{\boldmath $X$}})\times
(1+\lambda^2\Box){\mbox{\boldmath $B$}}  
\quad ,
\end{eqnarray} 
where the local fluid velocity is related to the electric current
by ${\mbox{\boldmath $v$}}=({c}/{q}{\rho_0}){\mbox{\boldmath $J$}}_e $.
When the vortex is almost at rest $(\dot{\mbox{\boldmath $X$}}\simeq 0) $
and the photon momentum is smaller than $1/\lambda$
 ($\lambda^2 \Box {\mbox{\boldmath $B$}} \simeq 0)$,
  the right hand side of (\ref{magL})
  reduces to the Lorentz force, 
$(1/c){\mbox{\boldmath $J$}}_e \times{\mbox{\boldmath $B$}}$ \cite{Kim}.

Now let us evaluate the effective action for 
vortices in charged superfluids by taking into account the
 phonon and photon fluctuations. 
Analogous to the neutral superfluid,  a stationary  
solution without vortex singularities  reads
\begin{eqnarray}
a_{\mu}=0 \quad ,\quad\rho=\rho_0 \quad ,\quad
\frac{\hbar}{m}\nabla\theta_{bg}
-\frac{q}{mc}{\mbox{\boldmath $A$}}_{bg}={\mbox{\boldmath $v$}}_{bg}
\quad {\rm and} \quad 
\dot{\theta}_{bg}
=-\frac{m}{2\hbar}{\mbox{\boldmath $v$}}_{bg}^2 \nonumber\\
 \partial^{\mu} F_{\mu 0,bg}=\frac{q}{c}\rho_0+\rho_{lat}=0\quad ,\quad
\partial^{\mu} F_{\mu i,bg}=-\frac{q}{c}\rho_0{v}_{bg}^i
\quad \quad .
\label{bgsc}
\end{eqnarray}
 Deep inside bulk superconductors in three dimensions,
 neither the magnetic field nor the 
electric current present.
 On the other hand, in the surface region
 of a bulk superconductor and in a 
 thin superconducting film,
 the electric current is non-vanishing
 due to the effect of the boundary conditions.
 These features are a consequence of the above 
 equations.
 In the following, we will mostly consider a thin 
 film placed in the
 $x-y$ plane and assume that 
 the transport current (the background velocity)
 is  constant in the $x-y$ direction.

By expanding $\rho (x)$, $\theta (x)$ and $A_{\mu}(x)$ 
in (\ref{TLGSC}) around the above  solution (\ref{bgsc}),
we get an effective action 
\begin{eqnarray}
{\cal Z}=\int {\cal D}
[\delta\rho , \delta \theta, \delta A_{\mu} ,a_{\mu},  b_{\mu\nu}]
e^{iS(\rho , \theta ,A_{\mu} , a_{\mu},  b_{\mu\nu};X)}
=e^{iS_{eff} (X)}\ \ ,
\label{efsc}
\end{eqnarray}
where
\begin{eqnarray}
{\cal  L}&=&{\cal L}_{st} 
+\delta\rho \left( \hbar a_0 -\hbar{\mbox{\boldmath $v$}}_{bg}\cdot
{\mbox{\boldmath $a$}} 
-\frac{\hbar^2}{2m}{\mbox{\boldmath $a$}}^2 \right)
+\frac{q\rho_0}{c}\delta {\mbox{\boldmath $A$}}\cdot
\frac{\hbar}{m}{\mbox{\boldmath $a$}}
\nonumber\\ && 
+\frac{1}{2}
(\delta\rho\quad\delta\theta\quad 
\delta A_0 \quad \delta{\mbox{\boldmath $A$}})
G^{-1}
\left( \begin{array}{c} 
        \delta\rho\\\delta\theta\\ 
\delta A_0\\ \delta{\mbox{\boldmath $A$}}
\end{array} \right)
-\delta\rho\hbar{\mbox{\boldmath $a$}}\cdot \frac{\hbar}{m}
(\nabla\delta\theta-\frac{q}{c\hbar}\delta {\mbox{\boldmath $A$}})
+O(\delta^3)
+\left( bf,\ bJ \right) \ \ ,
\nonumber  \\ 
{\cal L}_{st} &=&
\rho_0 \left( \hbar a_0 -
\hbar{\mbox{\boldmath $v$}}_{bg}\cdot{\mbox{\boldmath $a$}}
-\frac{\hbar^2}{2m}{\mbox{\boldmath $a$}}^2 \right)
 -\frac{1}{4}F_{bg,\mu\nu}F_{bg}^{\mu\nu}\ \ ,
\label{TLGC2}
\end{eqnarray}
with Hermitian matrix $G^{-1}$
\begin{eqnarray}
G^{-1}=\left(\begin{array}{cccc}
        \left( -2g+\frac{\hbar^2}{4m\rho_0}\triangle \right)& 
        -\hbar(\partial_0+{\mbox{\boldmath $v$}}_{bg}\cdot\nabla)&
        -q/c  &
        \frac{q}{c}{\mbox{\boldmath $v$}}_{bg}\\ &&& \\
        \hbar(\partial_0+{\mbox{\boldmath $v$}}_{bg}\cdot\nabla)&
        \frac{\rho_0\hbar^2}{m}\triangle&0&0\\ &&& \\
        -q/c &0&
        -\triangle & 
        0
        \\ &&& \\
        \frac{q}{c}{\mbox{\boldmath $v$}}_{bg}
        & 0&0&-(\Box +\frac{1}{\lambda^2} )
                \end{array}\right)
                \quad\quad .
\end{eqnarray}            
Here the Coulomb gauge for the $a$-field and 
 $\delta A$-field are used.
For zero background velocity, $G$ has poles
at
\begin{eqnarray}
\omega^2 = \frac{1}{\lambda^2} 
+{\mbox{\boldmath $k$}}^2 \ \ ,\quad\quad 
 \omega^2= \frac{1}{\lambda^2} 
+\epsilon^2({\mbox{\boldmath $k$}}^2)\quad\quad ,
\end{eqnarray}
where $\epsilon$ is the Bogoliubov spectrum (\ref{Bog}).
The massless Goldstone mode has been absorbed into
 massive mode. 
In order to take into account density fluctuations and the
photon effect, we integrate out these fluctuations;
\begin{eqnarray}
&&{\cal L}_{eff}(a_{\mu},b_{\mu\nu};X) \approx {\cal L}_{st}
\nonumber\\&&\quad\quad\quad
-\frac{\rho_0}{2m}
\left( \hbar (a_0-{\mbox{\boldmath $v$}}_{bg}
\cdot{\mbox{\boldmath $a$}})
-\frac{\hbar^2}{2m}{\mbox{\boldmath $a$}}^2
\right)
\frac{\triangle}{(\partial_0 
+{\mbox{\boldmath $v$}}_{bg}\cdot\nabla)^2+
\epsilon^2+1/\lambda ^2}
\left(\hbar (a_0-{\mbox{\boldmath $v$}}_{bg}
\cdot{\mbox{\boldmath $a$}})
-\frac{\hbar^2}{2m}{\mbox{\boldmath $a$}}^2\right)
\nonumber\\
 &&\quad\quad\quad
+\frac{\rho_0\hbar^2}{2m}
{\mbox{\boldmath $a$}}\cdot 
\frac{1/\lambda^2 }{\Box +1/\lambda^2 }
{\mbox{\boldmath $a$}} +O({\mbox{\boldmath $v$}}^2/c_s ^2)
\quad\quad .
\label{SCeff}
\end{eqnarray}

Suppose that a thin superconducting
  sample is set in the $x-y$ plane and the magnetic
field is applied in the $z$-direction. Vortices lie along the 
$z$ direction.
Suppose also that the transport current 
${\mbox{\boldmath $J$}}_e $ flows in the 
$x-y$ plane. By carrying out the 
integration of the $b$-field and using the solutions (\ref{hata}), 
the first term of the effective action (\ref{SCeff}) becomes
\begin{eqnarray}
&&-\frac{\rho_0\hbar^2}{2m}
\int d^2x \quad (a_0-{\mbox{\boldmath $v$}}_{bg}
\cdot{\mbox{\boldmath $a$}}) \left( 
\frac{\triangle}{(\partial_0 +{\mbox{\boldmath $v$}}_{bg}
\cdot\nabla)^2+
\epsilon^2+1/\lambda ^2}
\right) (a_0-{\mbox{\boldmath $v$}}_{bg}
\cdot{\mbox{\boldmath $a$}}) \nonumber\\
&&\quad\quad\quad\approx 
 \frac{1}{8\pi^2}\left( \frac{\phi_0 }{\lambda c_s }\right)^2
 (\sum_{a=b}+\sum_{a\neq b}) n_a n_b  
(\dot{\mbox{\boldmath $X$}}_a -{\mbox{\boldmath $v$}}_{bg})^i
(\dot{\mbox{\boldmath $X$}}_b -{\mbox{\boldmath $v$}}_{bg})^j
\epsilon^{ik}\epsilon^{jl}
\nonumber\\
&&\quad\quad\quad\quad
\frac{\triangle_X}{\triangle_X -1/(\lambda c_s)^2}
\int d^2x
\partial_k \ln |{\mbox{\boldmath $x$}}
-{\mbox{\boldmath $X$}}_a|\partial_l 
\ln |{\mbox{\boldmath $x$}}-{\mbox{\boldmath $X$}}_b|
\nonumber\\
&&\quad\quad\quad= 
\sum_a\frac{1}{2} m_{eff,a} 
(\dot{\mbox{\boldmath $X$}}_a-{\mbox{\boldmath $v$}}_{bg}) ^2
\nonumber\\
&&\quad\quad\quad\quad
 +\frac{1}{8\pi}\left( \frac{\phi_0 }{\lambda c_s }\right)^2
 \sum_{a\neq b} n_a n_b  
        \{(\dot{\mbox{\boldmath $X$}}_a-{\mbox{\boldmath $v$}}_{bg})
  \cdot(
        \dot{\mbox{\boldmath $X$}}_b -{\mbox{\boldmath $v$}}_{bg})
\left( K_0(\frac{|{\mbox{\boldmath $X$}}_{ab}|c}{\lambda c_s})
+\frac{1}{\triangle -1/(\lambda c_s)^2}\frac{2}{|X_{ab}|^2}
\right)
\nonumber\\
&&\quad\quad\quad\quad
-(\dot{\mbox{\boldmath $X$}}_a -{\mbox{\boldmath $v$}}_{bg}) \cdot 
\left(\frac{4}{\triangle -1/(\lambda c_s)^2}
\frac{  {\mbox{\boldmath $X$}}_{ab}{\mbox{\boldmath $X$}}_{ab}}
{ {\mbox{\boldmath $X$}}_{ab}^4} \right)
\cdot(\dot{\mbox{\boldmath $X$}}_b -{\mbox{\boldmath $v$}}_{bg}) \}
\quad\quad .
\label{a0a0X}
\end{eqnarray}
 Here $2\pi K_0 (x)=i\pi^2 H_0 (ix)$ with $H_0$ being 
a Hankel function of the first kind.
 The time derivative $(\partial_0-{\mbox{\boldmath $v$}}
\cdot\nabla)^2$
is neglected because 
$\dot{\mbox{\boldmath $X$}}$ and ${\mbox{\boldmath $v$}}$
 are much smaller than
$c_s$, and the Bogoliubov spectrum $\epsilon^2 $ is 
approximated by $-c_s ^2\triangle$ in the adiabatic approximation.
In the last line of (\ref{a0a0X})
the interaction part is neglected, because  
$|{\mbox{\boldmath $X$}}_{ab}| c>>\lambda c_s$.

One should note the qualitative difference between the
 kinetic term in the superfluid case  (\ref{easf})
 and that in the superconductor case  (\ref{a0a0X}).
 ``Center of vorticity'' enters into the kinietic term in the
 former, while
 each vortex has separate contributions to the kinetic term in
 the latter. In fact, in the superconductor,
 charge screening
 makes a vortex-size finite and
 each vortex can carry individual mobility and 
 energy. 
The inertial mass of a single vortex is estimated using 
the previous regularization 
with the coherence length $\xi$ as an ultraviolet cut off,
\begin{eqnarray}
m_{eff,a}&=&
 \frac{1}{4\pi}\left( \frac{\phi_0 }{\lambda c_s }\right)^2
  n_a ^2 K_0(\frac{\xi c}{\lambda c_s}) \ \ ,
  \nonumber \\ 
 &\approx&  \frac{1}{4\pi}\left( \frac{\phi_0 }{\lambda c_s }\right)^2
  n_a ^2 \left\{ \begin{array}{lcl}
        \ln \frac{\lambda c_s}{\xi c} &,&\ \ \xi c <<\lambda c_s \\
        \sqrt \frac{\lambda\pi c_s}{2\xi c} e^{-\xi c/\lambda c_s}
        &,&\ \ \xi c >>\lambda c_s \ \ .
         \end{array} \right.
\label{meffsc}
\end{eqnarray}
 So far, we have been adopting
 the   vorticity distribution with  the 
 $\delta$-function form  (\ref{eq.2}),
for simplicity. In this approximation,
  the ultraviolet cutoff
should be the coherence length $\xi$.
 For neutral superfluids, $\xi$
is nearly equal to the atomic scale $d$ which is the 
cutoff of the theory, so the $\delta$-function
 vorticity is a good approximation.
For superconductors, however, $\xi$ 
 is larger than the atomic scale, so the vortex-core contribution
 could  be non-negligible.
 In order to estimate the core contribution,
 let us separate
 the integral of the left hand side of  (\ref{a0a0X}) as
 $\int_{r=\xi}^{\infty} d^2x \rightarrow
\left(\int_{r=d}^{\xi} +\int_{r=\xi}^{\infty}\right) d^2x 
$:
 The second term, which is a contribution from  outside the
core, is given by (\ref{a0a0X}) and (\ref{meffsc}).
The first term from  inside the core, can be evaluated 
 in  the following way. 
In the vorticity tensor, 
the $\delta$-function is replaced by
 the regularized function, for example 
\begin{eqnarray}
\delta^{(4)} (x-X) \rightarrow 
\delta (x_0-X_0) \delta (x_3-X_3)
\left(\frac{1}{\sqrt{\pi}\xi}\right)^2
e^{-(\frac{{\mbox{\boldmath $x$}}
-{\mbox{\boldmath $X$}}}{\xi})^2} \ \ ,
\label{rgdlt}
\end{eqnarray}
 which is  introduced in \cite{SY}. 
So the vorticity vector ${\mbox{\boldmath $\omega$}}$
has a finite peak at $X$ with width $\xi$. 
 By solving the constraint (\ref{fdaJ}), one can show that  the
velocity around a vortex, 
${\mbox{\boldmath $a$}}$, has a maximum value 
at the edge of a core and
decreases toward a vortex center, then vanishes at the vortex center.  
So we replace $\nabla \ln  r$ in $a_{\mu}$
by $({\mbox{\boldmath $r$}}/\pi\xi^2)P(r/\xi)$ where 
$P$ is a solution of (\ref{fdaJ});
$\nabla\cdot({\mbox{\boldmath $r$}}P(r/\xi)/\pi\xi^2)
=e^{-(({\mbox{\boldmath $x$}}
-{\mbox{\boldmath $X$}})/\xi)^2}/(\pi\xi^2)$.
  One can also show that $a_{\mu}$ 
has linear dependence in ${\mbox{\boldmath $r$}}$
near the vortex center with $P(0)=1$.
Then the core contribution
becomes
\begin{eqnarray}
 &&\frac{1}{2}\left( \frac{\phi_0 }{\lambda c_s }\right)^2 n_a^2
 \int_{r=d}^{\xi} 
d^2x \quad \dot{\mbox{\boldmath $X$}}^2 \ 
\frac{{\mbox{\boldmath $r$}}P(r/\xi)}{\pi\xi^2} 
\cdot
\left( 
\frac{\triangle}{\triangle -1/(\lambda c_s)^2 }
\right) \frac{{\mbox{\boldmath $r$}}P(r/\xi)}{\pi\xi^2} 
\nonumber\\&&\quad\quad
 =\frac{1}{2}\left( \frac{\phi_0 }{\lambda c_s }\right)^2 n_a^2
\alpha(\frac{d}{\xi},\frac{1}{\lambda c_s})
\dot{\mbox{\boldmath $X$}}^2 ,
\end{eqnarray}
where
\begin{eqnarray}
\alpha(\Lambda ,\mu) 
\equiv
 \frac{1}{\pi^2}  \int_{\Lambda}^{1} 
d^2\hat{r} \quad  \hat{\mbox{\boldmath $r$}}P(\hat{r})
\cdot
\left( 
\frac{\hat{\triangle}}{\hat{\triangle} -\mu^2 }
\right) \hat{\mbox{\boldmath $r$}}P(\hat{r}) \quad\quad ,
\label{alpha}
\end{eqnarray}
 with $\hat{\mbox{\boldmath $r$}}\equiv {\mbox{\boldmath $r$}}/\xi$.
For a small core, $\xi \sim d$, $\alpha$ is almost zero, since
$\alpha(\Lambda ,\mu) \approx
 \frac{2}{\pi} (1-\Lambda)  
[\hat{r} \hat{\mbox{\boldmath $r$}}P(\hat{r})
\cdot
\left( 
\frac{\hat{\triangle}}{\hat{\triangle} -\mu^2 }
\right) \hat{\mbox{\boldmath $r$}}P(\hat{r}) ]|_{\hat{r}=(1-\Lambda)/2}
$.
For a large core, $\xi >>d$, $\alpha$ becomes a constant of order $1$,
since $\left( 
\frac{\hat{\triangle}}{\hat{\triangle} -(\xi c/\lambda c_s)^2 }
\right) $ takes 
$1$ for $\hat{\triangle} >>(\xi c/\lambda c_s)^2  $,
constant of order $1$ for $\hat{\triangle} \approx (\xi c/\lambda c_s)^2$,
and $0$ for $\hat{\triangle} <<(\xi c/\lambda c_s)^2  $.
The resultant effective mass 
 including with the contribution from outside the core is
\begin{eqnarray}
m^{tot}_{eff,a}=\frac{1}{4\pi}
\left( \frac{\phi_0 }{\lambda c_s }\right)^2 n_a^2
(K_0(\frac{\xi c}{\lambda c_s})
+\alpha (\frac{d}{\xi},\frac{1}{\lambda c_s})
) \quad\quad .
\label{core3}
\end{eqnarray}
 Our result is consistent with that in ref.\cite{DLg}
 for  $\xi c <<\lambda c_s$. However,
 our formula is not limited to this parameter region, while
 that of \cite{DLg} is only valid in this region.
 For conventional superconductors, we have the Landau parameter
$\lambda/\xi \approx 10$ and a sound velocity
  $c_s\approx 10^{-4}c$, while
 for high $T_c$ superconductors we have  
$\lambda/\xi \approx 10^3$ and $c_s\approx 10^{-3}c$. 
 Thus the core of vortices in conventional superconductors is 
relatively large, $d/\xi \sim 10^{-4}$,
$\alpha \sim 10^{-6}$ and $K_0\sim 10^{-400}$.
On the other hand, that in high $T_c$ superconductors is small,
 $d/\xi \sim 10^{0\sim-1}$, 
$\alpha \sim 0$ and $K_0\sim 0.4$. Therefore, 
the effective vortex mass in a conventional superconductor 
 mainly comes from inside a core
, while 
that of high $T_c$ superconductor mainly comes from outside 
the core.

Now let us return to the evaluation of the 
${\mbox{\boldmath $a$}}^2$ term in
 $S_{eff}(a_{\mu};X)$.
The second term of the effective action (\ref{SCeff}), 
when added to 
 the term in the stationary action, ${\cal L}_{st}$, becomes
\begin{eqnarray}
&&-\frac{\rho_0\hbar^2}{2m}\int d^2 x
 {\mbox{\boldmath $a$}}\cdot \left( 
1-\frac{1/\lambda^2 }{\Box +1/\lambda^2 }
\right) {\mbox{\boldmath $a$}}
\nonumber\\
&&\quad\quad\quad\quad\approx 
-\frac{1}{8\pi^2}\left( \frac{\phi_0}{\lambda }\right)^2
 (\sum_{a=b}+\sum_{a\neq b}) n_a n_b
\frac{\triangle_X}{\triangle_X -1/\lambda^2 }
\int d^2 x
\partial_i\ln|{\mbox{\boldmath $x$}}-{\mbox{\boldmath $X$}}_a|
\partial_i\ln|{\mbox{\boldmath $x$}}-{\mbox{\boldmath $X$}}_b|
  \nonumber \\
&&\quad\quad\quad\quad= 
  -\sum_a E_{0,a}-
\frac{1}{4\pi}\left( \frac{\phi_0}{\lambda }\right)^2
 \sum_{a\neq b} n_a n_b
  K_0(\frac{|{\mbox{\boldmath $X$}}_{ab}|}{\lambda})
\nonumber\\
&&\quad\quad\quad\quad \approx 
  -\sum_a E_{0,a}
   - \frac{1}{4\pi} \left( \frac{\phi_0}{\lambda }\right)^2
\sum_{a\neq b} n_a n_b
\left\{ \begin{array}{lcc}
        \ln \frac{\lambda }{|{\mbox{\boldmath $X$}}_{ab}|}& ,&
        |{\mbox{\boldmath $X$}}_{ab}| <<\lambda \\
\sqrt{\frac{2\pi\lambda}{|{\mbox{\boldmath $X$}}_{ab}|}}
e^{-|{\mbox{\boldmath $X$}}_{ab}|/\lambda}& ,&
         |{\mbox{\boldmath $X$}}_{ab}| >>\lambda 
 \end{array}\right.
\nonumber\\
\label{aiaiX}
\end{eqnarray}
where the vortex energy is estimated
 to be 
\begin{eqnarray}
E_{0,a}&=&
 \frac{1}{4\pi}\left( \frac{\phi_0}{\lambda }\right)^2
  n_a ^2 K_0(\frac{\xi}{\lambda })
  \nonumber \\ 
 &\approx& \frac{1}{4\pi}\left( \frac{\phi_0}{\lambda }\right)^2
  n_a ^2 \ln \frac{\lambda }{\xi} 
\quad ,\quad \xi <<\lambda \ \ .
\end{eqnarray}
The total vortex energy with the core contribution reads
\begin{eqnarray}
E^{tot}_{0,a}&=&
 \frac{1}{4\pi}\left( \frac{\phi_0}{\lambda }\right)^2
  n_a ^2 (K_0(\frac{\xi}{\lambda })
+\alpha(\frac{d}{\xi },\frac{1}{\lambda }))
  \nonumber \\ 
 &\approx& \frac{1}{4\pi}\left( \frac{\phi_0}{\lambda }\right)^2
  n_a ^2 \ln \frac{\lambda }{\xi} 
\quad ,\quad \xi <<\lambda \ \ .
\end{eqnarray}
The first part of the 
resultant $E_0^{tot}$ coincides with the free energy of 
 a vortex line per unit length given by  Abrikosov
\cite{Abr}.
The core contribution is generally  small;
for the conventional superconductor case 
$K_0\sim 3$ and $\alpha \sim 0.1$ and
for the high $T_c$ superconductor case 
$K_0\sim 1$ and $\alpha \sim 0$.

The vortex-vortex interaction
in (\ref{aiaiX}) has a natural form:
At small distances, the same logarithmic force
 as the superfluid vortices acts between the vortices.
 For large distances, the force is exponentially suppressed because of  
the electromagnetic shielding in superconductors.
Other vortex interactions
 are suppressed by the factor
${\mbox{\boldmath $v$}}^2/c_s ^2$.

The resultant effective action for vortices in a charged superfluid
 up to $O(\hbar^2)$ and $O((1/c_s)^0)$ is given by
\begin{eqnarray}
S_{eff}(X)=\int dt [\int d^2x \ {\cal L}_{em}
+ \sum_{a} \{{ L}_{Berry,a}+{ L}_{ve,a }+
{ L}_{kin,a}\}+{ L}_{int}]-U_{pin},
\nonumber\\ \label{easc}
\end{eqnarray}
\begin{eqnarray}
\left\{
\begin{array}{lcl}
{\cal L}_{em}
&=&
\frac{1}{2} ({\mbox{\boldmath $E$}}_{bg} ^2 -
{\mbox{\boldmath $B$}}_{bg} ^2)
\\ &&\\
{ L}_{Berry,a}&=&
\rho_0\pi\hbar n_a
{\mbox{\boldmath $e$}}_z
\cdot(\dot{\mbox{\boldmath $X$}}_a-
{\mbox{\boldmath $v$}}_{bg}) \times
({\mbox{\boldmath $X$}}_a-{\mbox{\boldmath $v$}}_{bg}t)
\\ &&\\
{ L}_{ve,a }&=&
- E^{tot}_{0,a}
\quad , \quad {\rm with}\quad
E^{tot}_{0,a}= \frac{1}{4\pi}\left( {\phi_0}/{\lambda }\right)^2
  n_a ^2 
   (K_0 (\frac{\xi}{\lambda })
+\alpha(\frac{d}{\xi},\frac{1}{\lambda}) )
        \\ && \\
{ L}_{kin,a}
&=&
\frac{1}{2}m^{tot}_{eff,a} 
(\dot{\mbox{\boldmath $X$}}_a-{\mbox{\boldmath $v$}}_{bg})^2
\quad , \quad{\rm with}\quad
m^{tot}_{eff,a}=\frac{1}{4\pi}
 \left( {\phi_0 }/{\lambda c_s }\right)^2
  n_a ^2 (K_0(\frac{\xi c}{\lambda c_s})
          +\alpha(\frac{d}{\xi},\frac{1}{\lambda c_s}) )
\\ && \\
{ L}_{int}
&=& 
   - \frac{1}{4\pi} \left( \frac{\phi_0}{\lambda }\right)^2
\sum_{a\neq b} n_a n_b K_0(\frac{|{\mbox{\boldmath $X$}}_{ab}|}
{\lambda})
 \end{array}\right. 
\nonumber
\end{eqnarray}
where ${\cal L}_{em}$ is the electromagnetic Lagrangian.

The equation of motion for an isolated  vortex is determined by
varying the effective action which is the sum of 
the acceleration term, 
the Magnus force  and  the pinning force.
A scenario for vortex motion is as follows:
 At first,  a vortex in pinned, 
so that $\dot{\mbox{\boldmath $X$}}$ is almost zero.
 If the transport current is large enough, 
the Magnus force becomes  strong  so that
the vortex starts to move by overcoming the pinning potential.
 A longitudinal voltage is thus produced.
 The Magnus force acts 
in such a way that the vortex velocity gets closer to
the transport velocity. 
Eventually the  vortex moves 
along with the transport current.
As a result,  Hall (transverse) voltage 
is produced.

\section{Summary and Concluding remarks}

We have presented a unified treatment of vortex dynamics 
at zero temperature
by using the topological Landau-Ginzburg theory. It is shown that 
the rank-two antisymmetric tensor potential, $b_{\mu\nu}$-field,
 is a fundamental field
coupled to vortex lines, such that the local coupling 
 ``$b_{\mu \nu} J^{\mu \nu}$" and the topological term 
``$\epsilon^{\mu\nu\rho\lambda}b_{\mu\nu}f_{\rho\lambda}$" 
 give rise to
the Magnus force in both neutral and charged superfluids.
Analogies with the electromagnetic theory have been shown;
the rank-two antisymmetric tensor potential, 
the hydrodynamical current and the Magnus force correspond to
the vector potential, the electromagnetic field and the 
Lorentz force, respectively.

 Although we treat $b_{\mu\nu}$ as a Lagrange multiplier in this paper,
 the action written in terms of $b_{\mu\nu}$ after 
 integrating out the $a_{\mu}$
 field in our theory (\ref{TLGSF})
 will have the same structure as the  Kalb-Ramond type
 theory \cite{KR}. In this case,
the $b_{\mu\nu}$-field becomes  dynamical. 
 This may clarify the relation between our theory and
 that in \cite{Zee,Dv} where $b_{\mu\nu}$ is treated as
 a dynamical field with a $H_{\mu\nu\lambda}^2$ term.

The phonon effect in a neutral superfluid 
was calculated and our main result is the 
effective action given by (\ref{easf}). 
In our  derivation,
``$\rho_0\hbar (a_0-{\mbox{\boldmath $v$}}_{bg}
\cdot{\mbox{\boldmath $a$}})$"
becomes
the Berry phase term after integrating out the $b_{\mu\nu}$-field
 and it leads naturally to  the Magnus force. 
 In deriving the Berry phase term and the Magnus force,
a first order formalism with single time-derivative is essential:
 Phenomenological 
 Landau-Ginzburg theory having second order time-derivative
 (such as that in \cite{DLg}) 
 cannot lead to the Berry phase term.

We provide a systematic method to regulate
vortex self-interactions such as the vortex energy and
the inertial  mass.  
 It turned out that  the vortex energy
and the inertial mass in  neutral superfluids
 can only be defined for the whole system and not for the
 individual vortices.
 In fact,  the vortex energy is proportional to the 
square of the total vorticity. Also, 
the  real dynamical degree of freedom is
``the center of vorticity'' of the system as is seen from the
 effective kinetic term.

 The
obtained  energy $E_0$ for an isolated vortex coincides 
with Feynman's chemical potential.
The obtained inertial mass 
 is about $m\rho_0 \gamma_0^2 /c_s^2$
 ($m$ is the atomic mass, $\rho_0$ is the condensation density,
 $\gamma_0$ is the unit vorticity and $c_s$ is
 the zero sound velocity).
They satisfy the relation; $m_{eff} =E_0/c_s^2$
which is consistent with the result in \cite{DLg}
where phenomenological Landau-Ginzburg theory is used
supplemented with the Fermi-liquid theory.
  We have clarified the
origin of the inertial mass by dividing fields 
into the stational configuration
  and the zero sound wave: 
 a vortex follows the classical
 hydrodynamical law, 
``$\dot{\mbox{\boldmath $X$}}
={\mbox{\boldmath $v$}}$'', if
  the zero sound wave is neglected.
  Once the zero sound wave is taken into account,
 a vortex gets an inertial mass and deviation from
  the classical hydrodynamics arises. Therefore 
  the zero sound wave should play an essential role
 for the pinning and depinning
phenomena and also for the quantum tunneling of vortices.

 An effective action with
  photon and density fluctuations 
  in charged superfluids is   given by (\ref{easc}). 
 The fundamental force acting on vortices 
in superconductors is also
the Magnus force. For a slowly moving magnetic flux (vortex),
 the magnus force  reduces to
  ``${\mbox{\boldmath $J$}}_e\times{\mbox{\boldmath $B$}}$" which 
  is the origin of the 
Lorentz force between the magnetic flux and the transport current.
 Vortices get  inertial 
mass through $\delta A_0$ and $\delta\rho$ fluctuations. 
 In contrast to the superfluid case,
 the vortex energy and the 
 inertial mass can be defined for individual vortices
 because of the charge screening in superconductors.
 Core contributions to the vortex energy and
 the inertial mass are also evaluated. 
 The obtained inertial mass is about
 $10^{-6}({1}/{4\pi})(\phi_0 /\lambda c_s)^2$ 
for conventional superconductor and
 $0.4({1}/{4\pi})(\phi_0 /\lambda c_s)^2$
 for high $T_c$ superconductor, where 
$\phi_0$ is the unit magnetic flux and $\lambda$ is the 
penetration depth.  
 Thus the main contribution to the effective inertial mass
 for conventional
superconductors comes from inside the core,
  while that for high $T_c$
superconductor comes from outside the core.
In reference \cite{DLg} opposite results are obtained 
for high $T_c$ superconductors. The reason for 
 this discrepancy comes from 
 the unjustified extrapolation of the formula
  in \cite{DLg} from the region $\xi c <<\lambda c_s$ to the
 region $\xi c\approx \lambda c_s$. We do not have such a problem,
 since our formula is valid for all parameter space.

Finally, we mention that the inertial mass of a vortex may be determined
 experimentally by the 
real-time observation of the vortex motion using  electron microscopy
\cite{Tono}.

\acknowledgements

We wish to thank P.Ao for fruitful discussions,
 critical comments and  collaboration in early stage of this work.
We are grateful to D.J.Thouless for instructive discussions.
We also wish to thank B.Hanlon for reading manuscript.
S.Y. was supported by 
Grant-in-Aid for Scientific Reserch from Ministry of Education,
Science and Culture (grant number 07854012).
M.S. is Fellow of the Japan Society for the Promotion of Science for 
Japanese Junior Scientists.
T. H. was supported by 
Grant-in-Aid for Scientific Reserch from Ministry of Education,
Science and Culture (grant number 06102004).

\newpage

\appendix
\section*{A}

In this appendix we derive the integral formula (\ref{rlt}):
\begin{eqnarray}
&&\int d^2x \partial_i
\ln|{\mbox{\boldmath $x$}}
-{\mbox{\boldmath $X$}}_a|\partial_j
\ln|{\mbox{\boldmath $x$}}-{\mbox{\boldmath $X$}}_b| 
\nonumber\\
 &&\quad\quad=
\left\{ \begin{array}{lcl}
-\pi[\delta^{ij}\{
\ln({|{\mbox{\boldmath $X$}}_{ab}|}/
{\varepsilon e^{1/2}})
-\ln({R}/{\varepsilon e^{1/2}})\}+
({X_{ab}^i X_{ab}^j}/{|{\mbox{\boldmath $X$}}_{ab}|^2}
-\delta^{ij}/{2})]
&,& {\rm for}{\mbox{\boldmath $X$}}_a\neq {\mbox{\boldmath $X$}}_b\\
\pi\delta^{ij}\ln({R}/{\varepsilon e^{1/2}})
&,& {\rm for}{\mbox{\boldmath $X$}}_a= {\mbox{\boldmath $X$}}_b
\end{array}\right.
\quad\quad .
\nonumber\\ &&
\label{rlta}
\end{eqnarray}
In the left hand side of (\ref{rlta}) there is an ultraviolet
 divergence 
at ${\mbox{\boldmath $X$}}_a={\mbox{\boldmath $X$}}_b$,
so 
we regulate $\ln |x|$ as 
$\ln\sqrt{x^2+\varepsilon^2}$ 
with a small parameter $\varepsilon$.
For an infrared divergence,  we  introduce
 a large cutoff parameter $R$ so that 
we neglect $1/R^2$ terms.
 Using Feynman's parameter formula
\begin{eqnarray}
\frac{1}{ab}=\int_0^1 dt \frac{1}{[at+b(1-t)]^2}\quad\quad ,
\end{eqnarray}
the left hand side of (\ref{rlta})  becomes
\begin{eqnarray}   
&&\int d^2x\frac{(x^i- X_a^i)(x^j- X_b^j)}
{(|{\mbox{\boldmath $x$}}
-{\mbox{\boldmath $X$}}_a|^2+\varepsilon^2)
(|{\mbox{\boldmath $x$}}-{\mbox{\boldmath $X$}}_b|^2+\varepsilon^2)}
\nonumber\\
&&\quad\quad
=\int^1_0 dt\int d^2x'
\frac{(x'^i+X_{ab}^i(t-1))(x'^j+ X_{ab}^jt)}
{[{\mbox{\boldmath $x$}}'^2
+t(1-t){\mbox{\boldmath $X$}}_{ab}^2+\varepsilon^2]^2}
\nonumber\\
&&\quad\quad
=-\pi[\delta^{ij}\{
-\frac{1}{2}+\ln \frac{\varepsilon}{R}
-\sqrt{\frac{1}{4}+\frac{\varepsilon^2}{|{\mbox{\boldmath $X$}}_{ab}|^2}}
\ln \left|\frac{1/2-\sqrt{1/4+ {\varepsilon^2}/
{|{\mbox{\boldmath $X$}}_{ab}|^2}}}
{1/2+\sqrt{1/4+ {\varepsilon^2}/
{|{\mbox{\boldmath $X$}}_{ab}|^2}}}\right|
 \}\nonumber\\
&&\quad\quad\quad
+\frac{X_{ab}^i X_{ab}^j}{|{\mbox{\boldmath $X$}}_{ab}|^2}
(1+\frac{\varepsilon^2}{|{\mbox{\boldmath $X$}}_{ab}|^2}
\frac{1}{\sqrt{{1}/{4}
+{\varepsilon^2}/{|{\mbox{\boldmath $X$}}_{ab}|^2}}}
\ln \left|\frac{1/2-
\sqrt{1/4+ {\varepsilon^2}/{|{\mbox{\boldmath $X$}}_{ab}|^2}}}
{1/2+\sqrt{1/4+ {\varepsilon^2}/
{|{\mbox{\boldmath $X$}}_{ab}|^2}}}\right|)
] \quad\quad .
\end{eqnarray}
It is easy to recover the right hand side of 
(\ref{rlta}) for both
${\mbox{\boldmath $X$}}_a \neq {\mbox{\boldmath $X$}}_b$ with  
$|{\mbox{\boldmath $X$}}_{ab}| >> \varepsilon$ and 
${\mbox{\boldmath $X$}}_a = {\mbox{\boldmath $X$}}_b$.
 Also, the above formula (A3)
 may be useful for a case where
 two vortices are close to each other
  $|{\mbox{\boldmath $X$}}_{ab}| \sim \varepsilon$.


\newpage


\vspace{0.5cm}
\begin{thebibliography}{99}
\bibliographystyle{unsrt}
%
\setlength{\itemsep}{0.0in}

\bibitem{Kim}  See for example, 
Y.B.Kim and M.J.Stephen, 
$Superconductivity$ edited. by R.D.Parks, (Dekker,New York,1969),\\
 F.G.de Gennes, $Superconductivity \ of\ metals\ and\ alloys$,
 (W.A.Benjamin, inc. 1965).
\bibitem{Till}
D.R.Tilley and J.Tilley, $Superfuidity\  and\ Superconductivity$
 ,(Adam Hilger, Bristol, 1990).
 


\bibitem{GPita}  V.T.Ginzburg and P.Pitaevskii, Sov. Phys. JETP
 {\bf 7} (1958) 858,\\
   P.Pitaevskii, Sov. Phys. JETP {\bf 13} (1961) 451,\\
  E.P.Gross Nuovo Cimento {\bf 20} (1961) 454.


\bibitem{JnOn}  T.W.Jing and N.P.Ong, 
Phys.\ Rev.\ {\bf B42} (1990) 10781,
and references cited therein.

\bibitem{WnTn}  S.J.Hagen et al.,
 Phys.\ Rev.\ {\bf B47} (1993) 1064,
and references cited therein.

\bibitem{AoThMag}  P.Ao and D.J.Thouless, Phys.\ Rev.\ Lett.
\ {\bf 70} (1993) 2158.

\bibitem{AT}  P.Ao and D.J.Thouless, 
Phys.\ Rev.\ Lett. \ {\bf 72} (1994) 132.

\bibitem{NAT}  Q.Niu, P.Ao and D.J.Thouless, 
Phys.\ Rev.\ Lett. \ {\bf 72} (1994) 1706.

\bibitem{TLG}  M.Hatsuda, S.Yahikozawa, P.Ao and D.J.Thouless, 
Phys.\ Rev. {\bf B49} (1994) 15870.

\bibitem{Gai}  F.Gaitan, Phys.\ Rev.\ {\bf B51} (1995) 9061,\\
J.\ Phys.\ Cond.\ Mat.\ {\bf 7} (1995) L165.

\bibitem{Klee}  K.Lee, cond-mat/9409046,
``Vortices and sound waves in superfluids".


\bibitem{Zee}  A. Zee, Nucl. Phys. {\bf B421} (1994) 111.
\bibitem{SY}  
M.Sato and S.Yahikozawa, Nucl. Phys. {\bf B436} (1995) 100.
\bibitem{Kura} H.Kuratsuji and H.Yabu, TMU-NT 940802,
``Magnus force for a single superfluid vortex:
 Hamiltonian dynamics approach based on the Ginzburg-Landau Lagrangian''.


\bibitem{Lamb}  H.Lamb, ``{Hydrodynamics}", (Cambridge University
press, 6ed. 1932) $\S 146$ and $\S 69$ .
 

\bibitem{Dor}  A.T.Dorsey, Phys.\ Rev.\ \ {\bf B46} (1992) 8376.

\bibitem{tunnel} A.O.Calderia and A.J.Leggett, Annals of Phys. {\bf 149}
 (1983) 374,\\
 G.Blatter, V.B.Geshkenbein and V.M.Vinokur,
Phys.\ Rev.\ Lett. \ {\bf 66} (1991) 3297,\\
 G.Blatter and V.B.Geshkenbein, Phys.\ Rev.\ \ {\bf B47} (1993) 2725,\\
M.J.Stephen, Phys.\ Rev.\ Lett. \ {\bf 72} (1994) 1534.



\bibitem{DLg}  J-M. Duan and A.J.Leggett,
 Phys.\ Rev.\ Lett. \ {\bf 68} (1992) 1216,\\
J-M. Duan, Phys.\ Rev.\ \ {\bf B48} (1993) 333.




\bibitem{Popov}  V.N.Popov, Theor. Mat. Fiz.,
 {\bf 6} (1971) 90 (65,English version),\\
Theor. Mat. Fiz., {\bf 11} (1972) 236 (478,English),
Theor. Mat. Fiz., {\bf 11} (1972) 354 (565,English);
 $Functional\ interals\ and\ collective\ excitaions$,
(Cambridge university press, 1987).

\bibitem{berry-p}  F.D.M.Haldane and Y.S.Wu, 
Phys.\ Rev.\ Lett. \ {\bf 55} (1985) 2887.

\bibitem{Fynm}  R.P.Feynman, chapter II in vol. I,  $Progress\
 in\ Low\ Temperature\ Physics$,  edited by
C.J.Gorter (1955). 

\bibitem{Abr}  Abrikosov,  Sov. Phys. JETP
 {\bf 5} (1957) 1174.

\bibitem{Dv}
R.L.Davis, Physica {\bf B178} (1992) 76.

\bibitem{KR} M.Kalb and P.Ramond, Phys.Rev. {\bf D9} (1974) 2273.

\bibitem{Tono} K.Harada et.al.,  Nature {\bf 360} (1992) 51.
 


 
\end{thebibliography}
\end{document}